\documentclass[11pt,twocolumn,pdflatex,sn-nature]{sn-jnl}
\setcitestyle{numbers,round}
\usepackage{geometry}
\usepackage{graphicx}
\usepackage{multirow}
\usepackage{amsmath,amssymb,amsfonts}
\usepackage{amsthm}
\usepackage{mathrsfs}
\usepackage[title]{appendix}
\usepackage{xcolor}
\usepackage{textcomp}
\usepackage{manyfoot}
\usepackage{booktabs}
\usepackage{algorithm}
\usepackage{algorithmicx}
\usepackage{algpseudocode}
\usepackage{listings}
\usepackage{xspace}
\usepackage{lipsum}
\usepackage{lineno}
\usepackage{microtype}         
\usepackage{caption}
\usepackage{titletoc}
\usepackage{overcite}

\DeclareCaptionLabelSeparator{pipe}{ \textbar\ }

\newcommand{\fmname}{\texttt{Tempov}\xspace}
\newcommand{\ed}{Extended Data\xspace}

\makeatletter
\AtBeginDocument{%
  \global\setlength{\baselineskip}{11.55pt}
}
\makeatother

\usepackage{titlesec}
\titlespacing*{\section}{0pt}{*0.9}{*0.3}  
\titleformat{\section}
  {\bf\normalsize}{}                       
  {0pt}{}

\begin{document}
\title[vs]{A satellite foundation model for improved wealth monitoring}

\author[1]{Zhuo Zheng}
\author[2]{Iván Higuera-Mendieta}
\author[3]{Richard Lee}
\author[4]{David Newhouse}
\author[4]{Talip Kilic}
\author[1]{Stefano Ermon}
\author[3,5]{Marshall Burke}
\author[2,3]{David B. Lobell}

\affil[1]{Department of Computer Science, Stanford University}
\affil[2]{Department of Earth System Science, Stanford University}
\affil[3]{Center on Food Security and the Environment, Stanford University}
\affil[4]{Development Economics Data Group, World Bank Group}
\affil[5]{Department of Environmental Social Science, Stanford University}

\abstract{
Poverty statistics guide social policy, but in many low- and middle-income countries, censuses and household surveys that collect these data are costly, infrequent, quickly outdated, and sometimes error-prone \cite{devarajan2013africa,seidler2025subnational}.
Satellite imagery offers global coverage and the possibility of predicting economic livelihoods at scale\cite{burke2021using,yeh2020using,chi2022microestimates}, yet existing approaches to predicting livelihoods with imagery or other non-traditional data often fail to reliably identify local-level variation and, as we show, degrade under temporal shift.
Here we introduce \fmname, a satellite foundation model pretrained by self-supervision on three million bi-temporal Landsat pairs and adapted with parameter-efficient fine-tuning to sparse survey labels.
The model enables large-scale, high-resolution wealth mapping and dynamic measurement, including zero-shot nowcasting up to a decade after observed labels, retrospective hindcasting, and decadal change tracking, while outperforming existing neural network and geospatial foundation-model baselines.
In low-label regimes, \fmname achieves competitive accuracy with only 10\% of survey samples, indicating substantially reduced dependence on expensive label collection.
The model further generalizes across populous countries within and outside Africa, and scales to a unified Africa-wide model with strong continent-level performance ($R^2=0.63$, $r^2=0.68$), from which we generate high-resolution decadal maps of wealth and wealth changes for the African continent. Analysis of these maps shows large variation in recent economic performance both within and across countries. 
Our open-source approach provides a pathway to timely, scalable, low-cost monitoring of wealth and poverty from routinely collected satellite data.
}

\maketitle

With more than 800 million people still living in extreme poverty globally, many sustainability challenges fundamentally stem from poverty and widening inequalities \cite{post2030v}.
Consequently, accurate and timely measurements of poverty and wealth are essential, providing the necessary data for the efficient allocation of scarce resources and forming the foundation for global efforts to monitor and improve human livelihoods \cite{jean2016combining}.
However, official poverty measurement in low and middle-income countries, primarily through household surveys, remains expensive, infrequent, and hard to scale. While combining surveys and census data can produce more precise small area estimates \cite{elbers2003micro,tarozzi2011can}, census data is also expensive, infrequent, and released with substantial temporal lags. The end result is significant spatial and temporal gaps in official statistics.
These deficiencies impede the objective assessment of progress toward sustainable development goals and constrain the targeting and evaluation of anti-poverty interventions \cite{devarajan2013africa,burke2021using}.

The expanding availability of remote sensing data, coupled with rapid advances in machine learning, has the potential to transform the measurement of socioeconomic conditions. 
Satellite imagery provides a scalable complement to traditional household surveys, offering consistent global coverage and visual proxies of development such as urban form, infrastructure, agricultural activity, and nighttime illumination \cite{engstrom2022poverty}.
Early studies primarily relied on nighttime lights to infer economic activity \cite{henderson2012measuring,chen2011using}. 
While effective at the country level, the coarse spatial resolution of these data (e.g., approximately 500\,m for the Visible Infrared Imaging Radiometer Suite) limits their ability to capture fine-scale heterogeneity, including neighborhood-level variation.
Moreover, nighttime lights are often dim and exhibit limited variation in the poorest regions, further constraining their usefulness \cite{jean2016combining,engstrom2022poverty}.
More recently, the use of high-resolution ($\leq$30 m) daytime satellite imagery combined with deep learning has enabled richer representations of local-level livelihoods at much finer spatial scales \cite{jean2016combining,babenko2017poverty,yeh2020using}.
As an alternative to deep learning–based methods, feature-engineering approaches have also proven useful. These methods typically use tree-based models such as XGBoost \cite{chen2016xgboost} with inputs comprised of various pre-defined (and thus more easily interpreted) features, including descriptors of mobile phone activity \cite{blumenstock2015predicting}, social media activity \cite{chi2022microestimates}, publicly available or proprietary geospatial covariates \cite{newhouse2024small},  and asset object counts derived from sub-meter–resolution commercial satellite imagery \cite{ayush2020generating}.

\begin{figure*}[ht]
\centering
\includegraphics[width=0.95\textwidth]{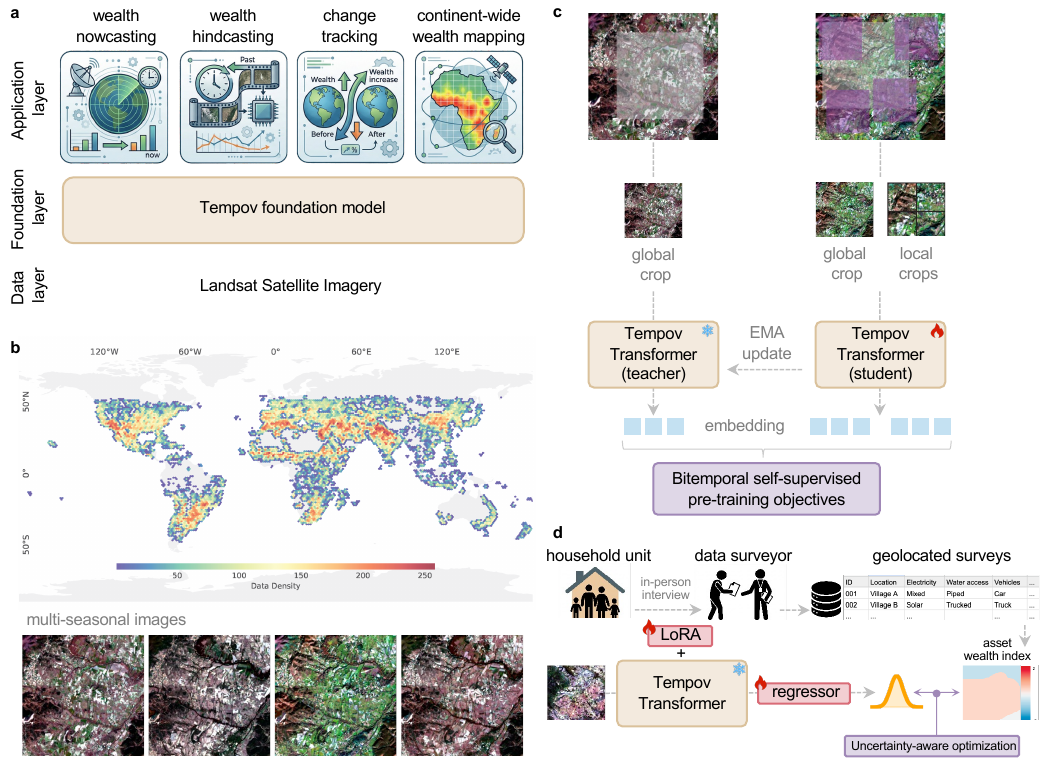}
\caption{\fmname framework for dynamic wealth measurement. 
\textbf{a}, \fmname models global satellite imagery and enables applications across dynamic wealth measurement, scaling from local villages to continent-wide mapping and spanning nowcasting, hindcasting, and change tracking.
\textbf{b}, Geographic distribution of the pre-training dataset. \fmname was pretrained globally on three-million bitemporal Landsat image pairs across 250k human settlement locations, capturing multi-seasonal phenological and radiometric variance. 
\textbf{c}, The bitemporal self-supervised pre-training paradigm, utilizing a teacher-student architecture with global and local crops to learn semantic and socioeconomic embeddings.
\textbf{d}, Pipeline for downstream wealth estimation tasks. 
The pre-trained \fmname model undergoes parameter-efficient fine-tuning and uncertainty-aware optimization to predict asset wealth index derived from geolocated household surveys.
}
\label{fig:1}
\end{figure*}

Despite these advances, three key limitations remain for large-scale dynamic wealth measurement. 
First, scarce ground-truth survey data hinder the end-to-end training of accurate deep poverty models \cite{yeh2020using,pettersson2023time,zheng2025dynamic}. 
Second, although auxiliary geospatial data can improve static wealth measurement \cite{chi2022microestimates,newhouse2025small}, their availability in most poor regions is typically very low, limiting the applicability of models that rely on these data. 
Third, even when such data are available, it is not clear that static features are well-suited for tracking \textit{changes} in wealth and poverty. 
Existing work shows relatively poor performance in change measurement on noisy available labels \cite{yeh2020using}. 
As a result, the potential of satellite-only models for large-scale dynamic wealth measurement remains underexplored.

\begin{figure*}[h]
\centering
\includegraphics[width=0.95\textwidth]{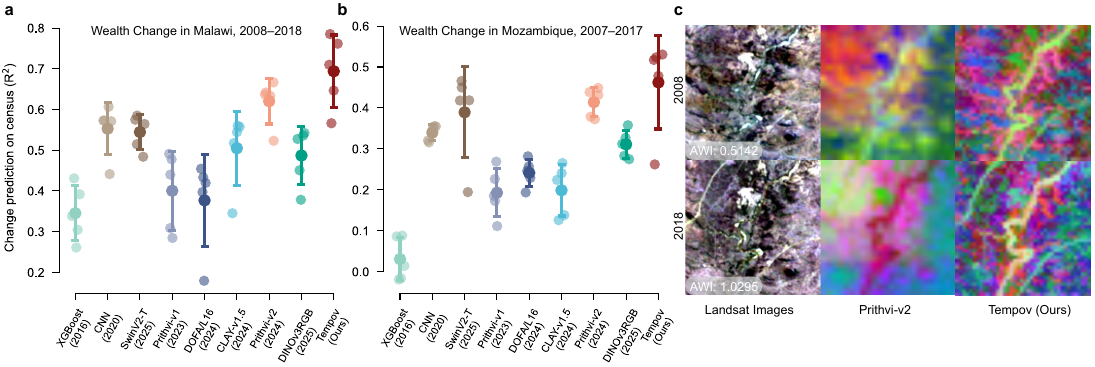}
\caption{Benchmarking wealth change prediction capabilities in data-rich scenario where the model is trained on full census data.
Performance comparison (coefficient of determination $R^2$) across Malawi (\textbf{a}) and Mozambique (\textbf{b}) for decadal tracking of wealth dynamics. 
The proposed method is evaluated against XGBoost, CNN, SwinV2-T and five geospatial foundation models. 
Data points represent independent validation folds ($n=5$); center dots and error bars indicate mean $\pm$ 1 s.d.
\textbf{c}, visualization of the learned semantic representation of two representative cases by showing the first three principal components of their high-dimensional embeddings.
}
\label{fig:2}
\end{figure*}

Here we introduce \fmname, a satellite foundation model designed to overcome the above challenges (Fig.~\ref{fig:1}).
The model consists of two parts: (1) an enhanced Vision Transformer (ViT) \cite{vit} designed to encode satellite images into high-dimensional embeddings; (2) a linear layer as a regressor to decode high-dimensional embeddings into socioeconomic predictions.
Instead of relying on generic visual features, \fmname inputs six atmospherically corrected surface reflectance bands (Blue, Green, Red, NIR, SWIR-1, and SWIR-2) from Landsat, selected to capture complementary biophysical attributes ranging from vegetation productivity and surface moisture to the structural characteristics of the built environment \cite{tucker1979red,small2004landsat,brown2008food,lobell2013use}, thereby providing robust physical proxies for socioeconomic estimation \cite{engstrom2022poverty}.
More model details can be found in Supplementary Information Section A.

While self-supervised pretraining enables feature extraction without human annotation \cite{cong2022satmae, simeoni2025dinov3}, applying it to wealth tracking faces a critical challenge: distinguishing genuine socioeconomic shifts from irrelevant seasonal fluctuations \cite{yeh2020using}. 
To address this, we curated a global, multi-seasonal dataset from SSL4EO-L \cite{stewart2023ssl4eo} spanning two decades (2001–2022) and constructed three million bi-temporal image pairs that maximize seasonal variance (Fig.~\ref{fig:1}b). 
We pretrained \fmname on this dataset using bitemporal self-supervised learning objectives (Fig.~\ref{fig:1}c; ``Methods''). 
This design forces the model to ignore transient radiometric noise while learning season-invariant semantics essential for tracking long-term changes in wealth.

Following pretraining, we adapt \fmname to a range of wealth estimation tasks by fine-tuning on household survey data (Fig.~\ref{fig:1}d).
Fine-tuning is substantially less costly than pretraining and typically requires only limited labeled data, enabling low-cost, privacy-preserving deployment by individual countries or institutes.
To account for the intrinsic uncertainty of household survey labels, we replaced the standard deterministic prediction with a probabilistic head that models outputs as a Gaussian distribution (``Methods''). 
This uncertainty-aware mechanism dynamically re-weights samples based on predicted variance, adaptively attenuating the impact of noisy labels during optimization.

\begin{figure*}[h]
\centering
\includegraphics{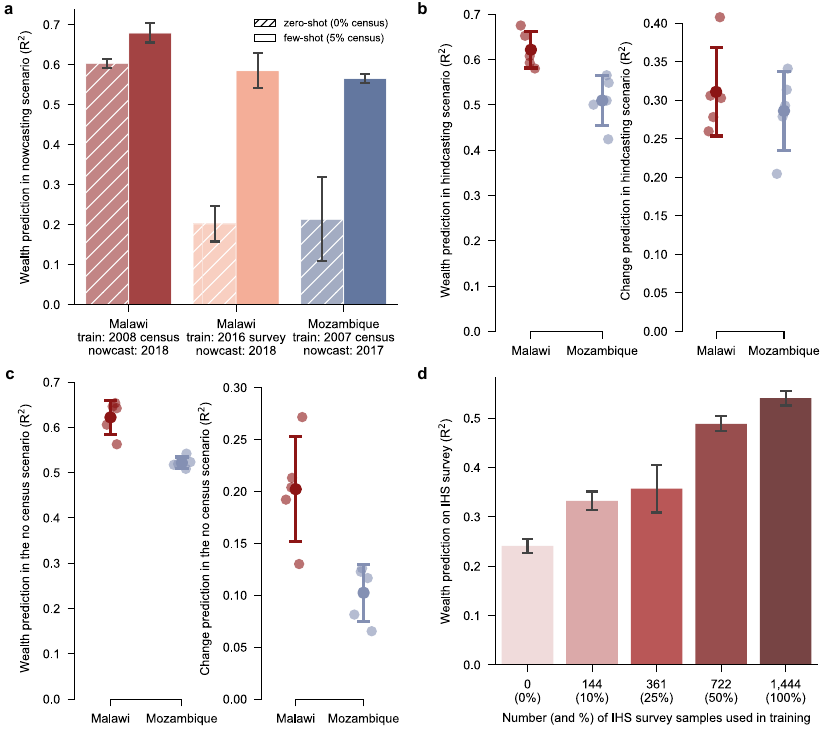}
\caption{Wealth measurement in data-scarce environments with \fmname.
\textbf{a}, Performance evaluation ($R^2$) of wealth nowcasting in Malawi and Mozambique. 
Models trained on historical data (Malawi 2008 census, Malawi 2016 survey, or Mozambique 2007 census) are evaluated on ground truth data from up to a decade later (2018 and 2017, respectively). 
The zero-shot setting (striped bars) evaluates the model directly on target-year imagery without fine-tuning, whereas the few-shot setting (solid bars) incorporates sparse target-year data (5\% census) for fine-tuning.
\textbf{b}, Retrospective hindcasting performance. 
Models are trained on most recent census (100\%) combined with sparse historical labels (5\%) to reconstruct past wealth states (left) and predict decadal wealth changes (right). 
\textbf{c}, Benchmarking under extreme data scarcity (no census, minimal survey). 
Wealth level and change prediction are evaluated when supervision is restricted to just 5\% of historical and 5\% of current census labels. 
\textbf{d}, Impact of increasing survey sample size on nowcasting accuracy. Models initially trained on the 2008 Malawi census are adapted using 2016 Malawi IHS survey data, with coverage ranging from 0\% to 100\%.
Data points denote $n = 5$ independent folds, and error bars in all panels indicate the mean $\pm$ 1 s.d.
Performance comparison against five geosaptial foundation models can be found in \ed Fig.~\ref{exfig:3} and \ed Fig.~\ref{exfig:4}a.
}\label{fig:3}
\end{figure*}

\section{Results}
\section{Static and dynamic wealth measurement}

\noindent
We first evaluated the model’s capability to estimate static asset wealth across two distinct socioeconomic contexts, Malawi and Mozambique, for which we have extensive and well-georeferenced ground data with repeated observations of the same locations over time.
Benchmarked against state-of-the-art geospatial foundation models, including DOFA \cite{xiong2024neural},  DINOv3 RGB (Sat) \cite{simeoni2025dinov3}, Prithvi v2 \cite{szwarcman2025prithvi} and CLAY-v1.5 \cite{clay_model_2024} and traditional supervised baselines, XGBoost, CNN \cite{yeh2020using} and ImageNet pretrained SwinV2-T \cite{zheng2025dynamic}, our framework consistently achieved superior predictive performance.
\fmname yields the highest coefficient of determination ($R^2$) in both 2008 and 2018 (average 87\% and 85\%) for Malawi (\ed Fig.~\ref{exfig:1}a, b) and 2007 and 2017 (average 74\% and 73\%) for Mozambique (\ed Fig.~\ref{exfig:1}c, d).

Accurate static estimation does not guarantee the ability to track changes over time, a task where traditional models often fail, perhaps due to overfitting to static geographic priors.
We tested this by predicting decadal changes in asset wealth, as measured by repeated wealth observations of the same locations in repeated decadal censuses.
While baseline models exhibited a sharp performance degradation in this dynamic setting, often dropping to near-random predictions, our model maintained robust predictive power (average 69\% for Malawi and 46\% for Mozambique, Fig.~\ref{fig:2}a, b).
This result confirms that our self-supervised pretraining strategy successfully disentangles genuine socioeconomic evolution from background environmental noise.

To understand the semantic basis of these predictions, we visualized the principal components of the learned embeddings (Fig.~\ref{fig:2}c).
The visualizations reveal that the model spontaneously attends to semantic features highly correlated with economic activity. 
For instance, the high-dimensional representations clearly delineate complex urban structures, road networks, and agricultural patterns from natural background, providing an interpretable basis for its wealth inference capabilities.
Furthermore, the feature embeddings generated by \fmname exhibit superior temporal consistency compared to those of other foundation models, such as Prithvi-v2. 
This stability is a key factor underlying \fmname's enhanced performance in estimating longitudinal wealth changes.

\begin{figure*}[h]
\centering
\includegraphics[width=0.95\textwidth]{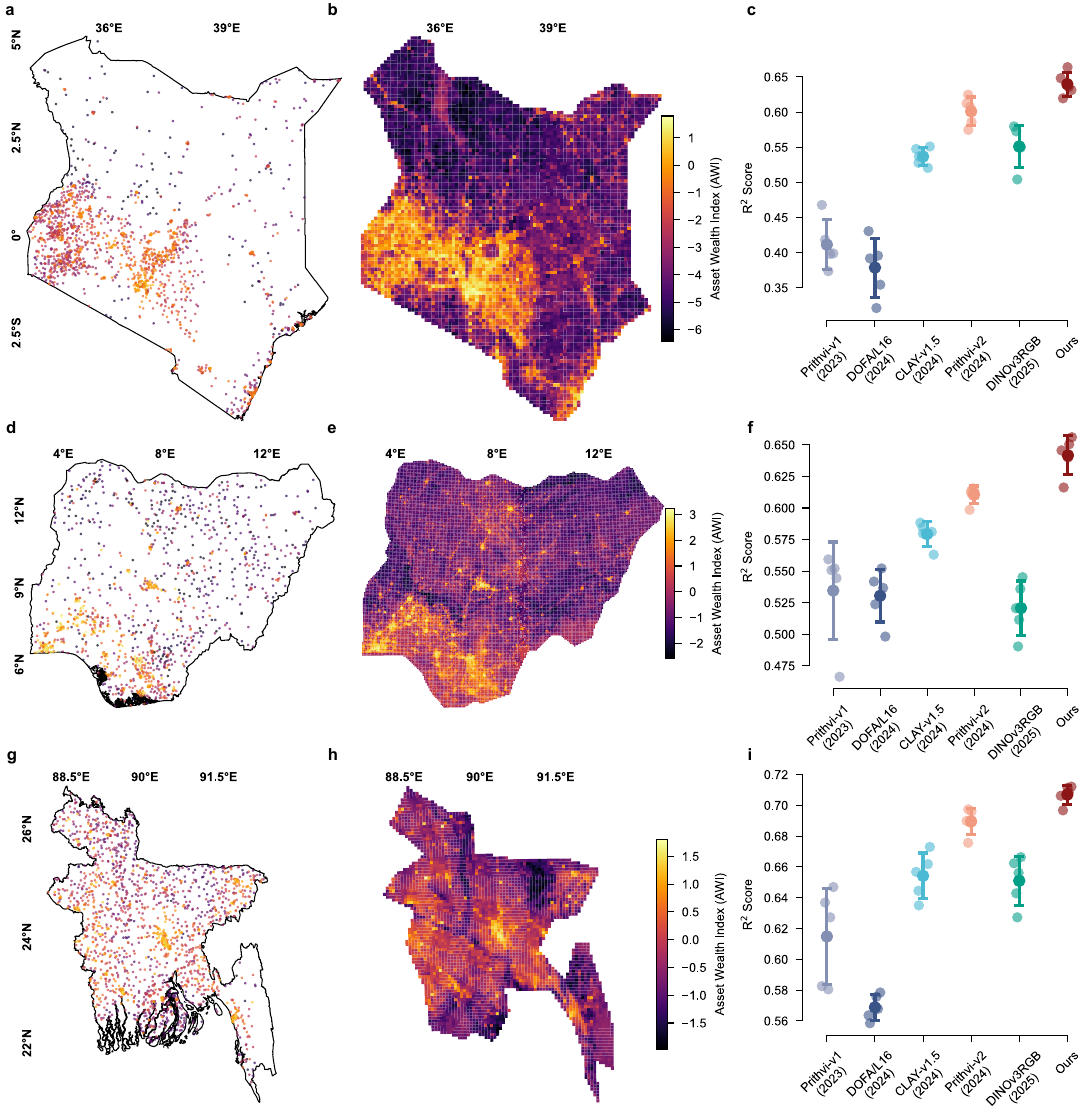}
\caption{The application of \fmname to DHS-based wealth mapping in Kenya, Nigeria, and Bangladesh.
\textbf{a, d, g}, Spatial distribution of the ground truth Asset Wealth Index (AWI) derived from Demographic and Health Surveys (DHS) clusters for Kenya (\textbf{a}), Nigeria (\textbf{d}), and Bangladesh (\textbf{g}).
\textbf{b, e, h}, Country-scale predicted AWI map for Kenya (\textbf{b}) at a spatial resolution of 10 km and for Nigeria (\textbf{e}) and Bangladesh (\textbf{h}) at 6 km. 
The predictions are generated by our model using median composites of Landsat imagery from 2020-2022 for Kenya, 2023-2025 for Nigeria, and 2016-2018 for Bangladesh.
\textbf{c, f, i}, Performance comparison of the proposed model against state-of-the-art geospatial foundation models in Kenya (\textbf{c}), Nigeria (\textbf{f}), and Bangladesh (\textbf{i}), evaluated using the $R^2$ score on the test set.
Error bars indicate the standard deviation across five runs.
}\label{fig:4}
\end{figure*}

\begin{figure*}[h]
\centering
\includegraphics[width=0.95\textwidth]{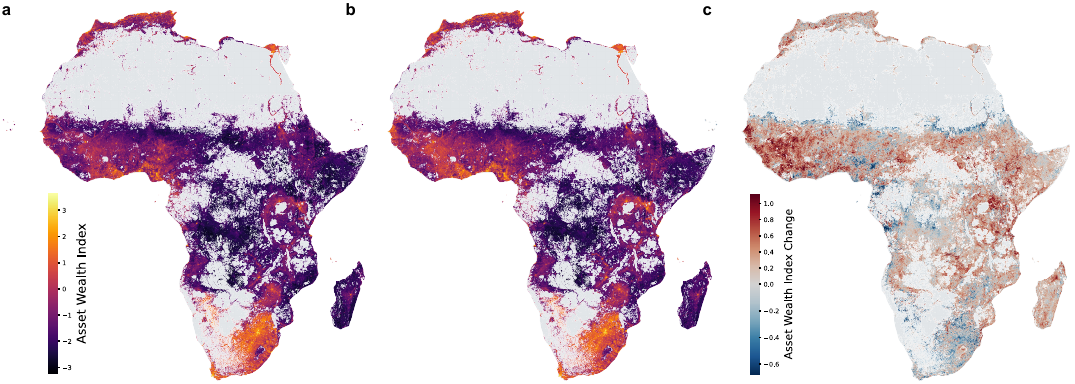}
\caption{High-resolution continent-wide maps of wealth and wealth change across Africa generated by Tempov.
\textbf{a}, Predicted wealth in 2015.
\textbf{b}, Predicted wealth in 2025.
\textbf{c}, Predicted decadal wealth change from 2015 to 2025. 
All maps are generated using a five-fold ensemble of Tempov models trained on all countries with available survey.
All predictions are wall-to-wall across Africa, but grid cells with a population density of 1 person per km$^{2}$ or less are masked in gray.
}\label{fig:5}
\end{figure*}

\section{Wealth measurement in data-scarce environments}

\noindent
A fundamental limitation to continuous poverty monitoring at granular geographic levels is the extreme scarcity of repeated observations of the same locations over time.
Available censuses are infrequent and quickly become outdated: we compute that the $R^2$ between asset wealth measures in earlier and later censuses is 0.42 in Malawi and -0.71 in Mozambique.
Meanwhile, cross-sectional surveys, such as from the Demographic and Health Surveys, are spatially sparse and rarely revisit identical locations \cite{yeh2020using}.
To address this practical setting, we use a two-stage adaptation strategy  (\ed Fig.~\ref{exfig:3}a) in which \fmname is first calibrated on historical census labels and then rapidly adapted to the target year with low-rank adaptation (LoRA) \cite{hu2022lora} using a smaller contemporary sample.

Fig.~\ref{fig:3}a evaluates this setting through decade-scale nowcasting in Malawi and Mozambique.
Under strict zero-shot transfer (i.e. training on one year and predicting on another; hatched bars), \fmname trained on historical labels retains substantial predictive power on target-year imagery, achieving state-of-the-art performance (\ed Fig.~\ref{exfig:3}b, c, d) and even matching or exceeding the few-shot performance of several geospatial foundation models (e.g., CLAY-v1.5, DOFA/L16; \ed Fig.~\ref{exfig:3}b).
When training data are limited (e.g., training on the sparse Malawi 2016 survey), zero-shot performance naturally declines.
However, by incorporating just 5\% of target-year data (few-shot, solid bars), the model's performance rapidly recovers (\ed Fig.~\ref{exfig:3}b, c, d). Notably, across settings with both dense and sparse training data, this zero-shot resilience and few-shot adaptability significantly outperform state-of-the-art geospatial foundation models (\ed Fig.~\ref{exfig:3}b, c, d), some of which suffer from severe negative transfer under such temporal shifts.

Beyond forward prediction, a robust foundation model should also enable ``retrospective hindcasting'', reconstructing historical wealth dynamics.
We test this by training the model on current data and adapting it to predict historical wealth using only sparse historical labels (5\%) (Fig.~\ref{fig:3}b).
\fmname reconstructs the past wealth distribution ($R^2 = 0.62$ and $0.51$ for Malawi 2008 and Mozambique 2007, respectively) and recovers decadal wealth change ($R^2 = 0.31$ and $0.29$ for Malawi and Mozambique, respectively) more accurately than competing methods (\ed Fig.~\ref{exfig:3}e), although with lower performance than if the full census in the earlier year was available for training (Fig.~\ref{fig:2}).
This bidirectional adaptability confirms that our model learns relevant temporally-varying socioeconomic semantics rather than overfitting to a specific timeframe.

To mimic the most resource-constrained deployment scenario, we simulate a ``no census'' regime , in which supervision is limited to small subsets of both historical (5\%) and current (5\%) labels.
Even in this ``no-census'' regime, \fmname preserves reasonable accuracy for both wealth level estimation and change prediction, whereas several alternatives deteriorate sharply (with $R^2$ dropping near zero; \ed Fig.~\ref{exfig:3}f).
This resilience suggests that our approach can help enable low-cost, high-frequency wealth monitoring even for regions lacking both comprehensive historical census baselines and extensive current surveys.

Finally, we systematically quantified the cost-efficiency of combining historical census priors with contemporary surveys. By scaling the available 2016 Malawi survey samples from 0\% to 100\% on top of a 2008 census prior, we observe a monotonic increase in nowcasting accuracy, with the steepest gains in low-label regimes (Fig.~\ref{fig:3}d).
Consistent with this trend, \ed Fig.~\ref{exfig:4}a shows that \fmname attains competitive performance with only 10\% of survey samples (144 image-wealth pairs), comparable to the level that baseline foundation models reach only with full (100\%) survey supervision.
Together, these results suggest that, in order to make accurate contemporary measurements, strong historical priors can offset the need for exhaustive contemporary sampling, provided that adaptation is anchored by a strong foundation model with socioeconomic representations.

\section{Scalability to standardized DHS surveys}

\noindent
Our previous evaluations established that historical census data provides critical structural priors for adapting to sparse surveys. 
However, a true test of global scalability requires generalizing to new geographies where such census data might not be available for training or evaluation.
To demonstrate this broader applicability of our framework, we extended our evaluation to three highly populous and distinct nations: Kenya (East Africa), Nigeria (West Africa), and Bangladesh (South Asia), leveraging village-level wealth data from the widely adopted Demographic and Health Surveys (DHS) as ground truth.
Unlike dense census tracts, DHS data, while globally available across over 90 countries, are constrained by their spatially sparse, cluster-based sampling design. 
Validating our foundation model in these settings is an important test of the model's capacity to scale across the geographically and socioeconomically diverse regions of the Global South.

Despite the relative sparsity of ground labels (Fig.~\ref{fig:4}a, d, g), fine-tuning our model on a subset of labels and evaluating on held-out labels exhibited superior generalization capabilities in all three countries, achieving the highest $R^2$ (average 0.66) among state-of-the-art geospatial foundation models, including Prithvi-v2 and Clay-v1.5 (Fig.~\ref{fig:4}c, f, i).
Notably, this superior performance was maintained across different target spatial resolutions (10 km for Kenya, 6 km for Nigeria and Bangladesh) and timeframes (2020–2022, 2023–2025, and 2016-2018), confirming that the learned spatiotemporal representations of \fmname are spatially and temporally adaptable and not artifacts of specific regional datasets.

The value of \fmname 's scalability is visualized in the transformation from sparse survey points to continuous, national-scale wealth maps (Fig.~\ref{fig:4}b, e, h).
The predicted maps accurately delineate known socioeconomic patterns: in Kenya, capturing the sharp contrast between the affluent central highlands and the economically marginalized arid northern regions;
in Nigeria, clearly identifying the prosperous southern coastal urban corridors (e.g., the Lagos megacity region) against the less wealthy northern interior; 
and in Bangladesh, successfully highlighting the dense concentration of wealth around the central capital region (Dhaka) and major urban hubs, contrasting with the broader, less affluent riverine peripheries.

\section{Continent-wide wealth maps for Africa}

The consistent performance of \fmname across space and time with minimal training data, along with the computational efficiency of the LoRA fine-tuning procedure, opens up the possibility for low-cost, high-resolution mapping of wealth at very large spatial scales.
To demonstrate viability, we first evaluated \fmname's spatiotemporal robustness across five distinct generalization strategies before deployment, with strategies reflecting whether data from inside and/or outside the target country/year was used to train the model (see Methods) (\ed Fig.~\ref{exfig:5}). 
The results reveal that a unified, all-African \fmname model trained on harmonized multi-country and multi-year DHS data (``All countries, in year'') achieves strong predictive accuracy ($R^2 = 0.63, r^2=0.68$), matching the performance of specialized, single-country models (``In country, in year'', $R^2 = 0.64, r^2=0.68$).  
Furthermore, even under the most stringent zero-shot geographical transfer setting (``Out of country''), the model maintains a high Pearson correlation ($r^2 > 0.65$), indicating that while absolute baseline calibration presents cross-border challenges, the underlying structural representations extracted by \fmname preserve the relative rank-ordering of wealth across unseen territories.

Given these robust generalization metrics, we trained five \fmname models under a five-fold cross-validation scheme (``All countries, in year'') using data from all 34 African countries with recent DHS surveys, and averaged their predictions to generate continuous 6 km $\times$ 6 km maps of asset wealth and wealth changes across the entire African continent for 2015 and 2025 (Fig.~\ref{fig:5}). 
Two static wealth maps (Fig.~\ref{fig:5}a,b) indicate substantial variation in wealth both across and within countries, with decompositions of the Theil index indicating that roughly 80\% of wealth inequality is within-country in both time periods.

The decadal wealth change map (Fig.~\ref{fig:5}c) again shows substantial variation in wealth gains across and within countries, with gains concentrated mainly in regions of West and East Africa, and substantial declines observed in parts of Southern and Central Africa.
These spatial patterns of wealth change broadly align with national-level time trends in GDP per capita for selected African countries (see Supplementary Information Section D).
Patterns of wealth changes show modest evidence of ``convergence", with initially lower-wealth areas growing only slightly faster on average than initially wealthier regions, consistent with recent cross-country national accounts data \cite{patel2025convergence} (\ed Fig \ref{exfig:6}a). 
Country-level factors also explain only about one-third of the variation in observed wealth changes (\ed Fig \ref{exfig:6}c), arguing against the simplest stories about the primacy of such factors (e.g. institutional quality) in explaining variation in economic performance.  Instead, we find that local-level trends in temperature and variation in the number of nearby conflict events are stronger predictors of wealth changes over the period (\ed Fig \ref{exfig:6}b). We emphasize that these estimates do not necessarily have a causal interpretation.

\section{Discussion}\label{sec:discuss}

\noindent
This study shows that a satellite foundation model can unify cross-sectional wealth estimation and decadal change tracking within a single framework.
\fmname generalizes to wealth prediction across diverse geographies, time periods, and survey densities, indicating that the learned representation transfers beyond country-specific training contexts.
Most importantly, this generalization enables a low cost, contiguous, high-resolution wealth map, which we demonstrate for the African continent for multiple years, including very recent years, produced with a unified model and consistent methodology.
To our knowledge, this is the first open-source, continent-wide, up-to-date African wealth map at this spatial granularity, and it demonstrates how strong foundation-model priors can turn sparse survey signals into policy-relevant population-scale monitoring products.

Several avenues may further improve \fmname.
First, prior studies show that auxiliary signals such as mobile-phone metadata, social-connectivity proxies, and object-level features from very-high-resolution imagery can improve socioeconomic prediction beyond satellite imagery alone \cite{blumenstock2015predicting,chi2022microestimates,ayush2020generating,newhouse2024small}. 
Extending \fmname with multimodal, missing-data-tolerant fusion could therefore increase accuracy where such covariates are available.
Second, work on multi-temporal imagery indicates that explicit temporal modeling can strengthen neighborhood-level poverty estimation and dynamic tracking \cite{pettersson2023time,zheng2025dynamic} 
Extending \fmname with longer time series and calibration may further improve out-of-time robustness.
Third, the small-area estimation literature emphasizes design-consistent calibration and uncertainty estimation \cite{elbers2003micro,tarozzi2011can,newhouse2024small}. 
Finally, it would be useful to empirically test the ability of foundational models such as \fmname to predict socioeconomic indicators beyond wealth, such as the consumption or income data typically used to construct official poverty statistics; such tests are currently constrained by the extreme paucity of geo-referenced public micro-data on these outcomes.

Beyond methodological gains, \fmname has the potential to reshape how evidence on economic livelihoods is produced and used, particularly in an era with declining resources devoted to DHS and other survey-based approaches.
By combining global pretraining with lightweight country-level adaptation, the framework can move wealth measurement from infrequent survey snapshots toward high-frequency, spatially detailed monitoring that supports subnational targeting, disaster response, and program evaluation.
Its open-source and continent-scale design also provides a transparent analytical backbone for cross-country comparability, reproducible benchmarking, and local customization by national statistical offices and development partners.

\bibliography{arxiv}

\clearpage

\section{Methods}\label{sec:method}

\section{Pretraining satellite imagery}

To enable \fmname to learn temporally invariant representations that are robust to phenological variation, we curated a large-scale bi-temporal Landsat dataset from the SSL4EO-L archive \cite{stewart2023ssl4eo}. 
The dataset comprises Landsat 7, 8 and 9 imagery collected between 2001 and 2022, spanning approximately 250,000 distinct human-settlement locations worldwide. 
For each location, we constructed bi-temporal image pairs from different seasons within the same year to maximize radiometric variation induced by vegetation phenology and illumination changes, yielding three million unlabeled pairs for pretraining.
We used six atmospherically corrected surface-reflectance bands (Blue, Green, Red, NIR, SWIR-1 and SWIR-2), selected for cross-sensor consistency and complementary sensitivity to built-up structure, vegetation status and surface moisture relevant to socioeconomic inference.

\section{Household survey data}

For downstream fine-tuning and evaluation, we compile asset-based wealth data from two primary sources:
(1) Census data: Full-count Population and Housing Censuses from Malawi (2008, 2018) and Mozambique (2007, 2017). 
These datasets provide spatially dense, enumeration-area-level aggregates of the Asset Wealth Index (AWI), computed via Principal Component Analysis (PCA) of household asset ownership (e.g., radio, bicycle, floor material) \cite{filmer2001estimating,sahn2003exploring}.
(2) Survey data: We use nationally representative surveys, including the Integrated Household Survey (IHS) for Malawi (2016) and Demographic and Health Surveys (DHS) covering nearly all African countries (89 country-year entries).
Fig.~\ref{fig:4} uses only DHS data from Kenya (2020, 2022), Nigeria (2024), and Bangladesh (2011, 2014, 2017), whereas Fig.~\ref{fig:5} uses DHS data from nearly all African countries (34 countries; see Supplementary Information Section B) for continent-wide wealth mapping.
DHS clusters are georeferenced with random displacement (jitter) of up to 2 km in rural areas and 5 km in urban areas for confidentiality. 
Following the wealth calculation used for the census data, we construct a comparable PCA-based wealth index representing average household wealth at the cluster level.

To achieve spatial and temporal consistency in categorical variables across DHS surveys, we apply the household-characteristic harmonization approach in \cite{colston2024spatial}, using the following ordinal categories: \textit{natural}, \textit{rudimentary}, and \textit{finished} for floor (\texttt{hv2013}), wall (\texttt{hv214}), and roof (\texttt{hv2014}) materials; \textit{open}, \textit{unimproved}, and \textit{improved} for toilet (\texttt{hv205}); and \textit{surface}, \textit{unimproved}, and \textit{improved} for drinking water source (\texttt{hv201}). 
We include only assets (e.g., motorcycle, mobile phone, clock) with no more than 10\% missing values across surveys over time (2010--2024). 
We show that our DHS AWI is robust to alternative selections of binary assets and remains highly correlated ($r \geq 0.95$) with an alternative index based on per-household asset counts and different asset compositions.

\section{Growth determinants data}
To validate the \fmname inference results, we examine the relationship between the model's wealth predictions and established growth determinants, as shown in \ed Fig.~\ref{exfig:6}. 
We focus on four key drivers: democratic institutional quality, remoteness, conflict, and environmental conditions. 
To measure democratic institutional quality, we compute the average Polity V score for each country between 2015 and 2020, and assign this value to all grid cells within the country. The Polity V index captures regime characteristics on a continuous scale ranging from $-10$ (fully autocratic) to $+10$ (fully democratic)~\cite{MarshallGurr2020}. 
We proxy remoteness using the distance from each grid cell's centroid to the national capital. Conflict intensity is measured by aggregating the total number of casualties from the Uppsala Conflict Data Program (UCDP) geo-referenced events, resampled to a 50-km resolution and summed over the period 2015--2024~\cite{SundbergMelander2013}. Finally, to capture environmental determinants of growth, we compute total temperature decade change for each grid using a linear trend between 2015 and 2025 using the ECMWF ERA5 reanalysis~\cite{Hersbach2020}.

\section{The Tempov model}
\fmname adopts an improved variant of the Vision Transformer (ViT-Large) architecture \cite{vit} (see implementation details in Supplementary Information Section A), specifically adapted for multispectral image pretraining.
To accommodate 6-channel satellite imagery (Blue, Green, Red, NIR, SWIR-1, SWIR-2) while leveraging rich feature representations learned from large-scale RGB pre-training (e.g., ImageNet or standard remote-sensing datasets), we modify the patch embedding layer.
Let $W \in \mathbb{R}^{D \times C_{\text{in}} \times K \times K}$ denote the weights of the patch projection layer, where $C_{\text{in}}=6$ and kernel size $K=16$.
We initialize the weights corresponding to the RGB channels ($C_{\text{RGB}}=\{\text{Blue, Green, Red}\}$) using pre-trained parameters, while the weights for the additional multispectral channels ($C_{\text{extra}} = \{\text{NIR, SWIR-1, SWIR-2}\}$) are initialized to zero:
\begin{equation}
W_{:,c,:,:}=
\begin{cases}
W^{\text{pre}}_{:,c,:,:}, & c\in C_{\text{RGB}}, \\
\mathbf{0}, & c\in C_{\text{extra}}.
\end{cases}
\end{equation}
This zero-initialization strategy ensures that, at the onset of training, contributions from the extra spectral bands are null, thereby preserving the original pre-trained feature distribution. 
During optimization, the model gradually learns to integrate information from these additional bands to capture biophysical surface properties \cite{lobell2013use}, such as vegetation health and moisture content that are critical for accurate socioeconomic inference.

For pre-training, \fmname takes a multispectral satellite image as input and outputs both a global embedding and a dense embedding for subsequent self-supervised training.
For fine-tuning and inference, we retain only the global embedding, which is fed into a linear layer to predict AWI.

\section{Training methods}
The overall training recipe consists of two stages: (1) bitemporal self-supervised pre-training and (2) fine-tuning with parameter-efficient, uncertainty-aware optimization.

\noindent\textbf{Bitemporal self-supervised pretraining objectives}.
To learn temporally invariant representations, we design a bitemporal variant of discriminative self-supervised learning (SSL) objectives \cite{simeoni2025dinov3}, which is based on image-level DINO loss \cite{caron2021emerging} and patch-level iBOT loss \cite{zhou2021ibot}.
The main idea is to sample two different seasonal views of Earth's surface instead of different geometric views of the same image in generic-purpose SSL \cite{caron2021emerging,ijepa}. 
The architecture consists of a student \fmname $g_{\theta_s}$ and a teacher \fmname $g_{\theta_t}$ (an exponential moving average of the student with a momentum of 0.992).
Given a bitemporal image pair $(x_1, x_2)$, where $x_t \in \mathbb{R}^{H\times W\times C}$ and $t\in\{1,2\}$, we denote one global crop by
\begin{equation}
x_t^{g} \in \mathbb{R}^{224\times224\times C},
\end{equation}
and four local crops by
\begin{equation}
x_{t,j}^{\ell} \in \mathbb{R}^{96\times96\times C}, \qquad j=1,\dots,4.
\end{equation}
We denote the local view set as $\mathcal{V}_t^{\ell}=\{x_{t,1}^{\ell},\dots,x_{t,4}^{\ell}\}$.
Thus, our bitemporal DINO (bi-DINO) loss can be written as:
\begin{equation}
\mathcal{L}_{\text{bi-DINO}}
= - \frac{1}{|\mathcal{V}_s|}\sum_{v \in \mathcal{V}_s} p_{\theta_t}(x_1^{g}) \log p_{\theta_s}(v).
\end{equation}
where $\mathcal{V}_s = \{x_2^{g}\} \cup \mathcal{V}_t^{\ell}$.
Here, $p_{\theta}(\cdot)=h^{\text{DINO}}_{\theta}(g_{\theta}(\cdot))$ denotes the class-token probability distribution produced by \fmname backbone $g_{\theta}$ followed by a DINO head $h^{\text{DINO}}_{\theta}$ \cite{caron2021emerging}.
Accordingly, $\mathcal{L}_{\text{bi-DINO}}$ is the cross-entropy loss from the teacher prediction on $x_1^g$ to the student predictions over all target views in $\mathcal{V}_s$, enforcing temporal consistency across seasonal views.
Simultaneously, for the patch-level objective, we employ a bitemporal iBOT (bi-iBOT) loss together with uniformity regularization \cite{sablayrolles2018spreading}.
\begin{equation}
\mathcal{L}_{\text{bi-iBOT}}
= - p^{\prime}_{\theta_t}(x_1^{g}) \log p^{\prime}_{\theta_s}(x_2^{g}).
\end{equation}
where $p^{\prime}_{\theta}(\cdot)=h^{\text{iBOT}}_{\theta}(g_{\theta}(\cdot))$ denotes the patch-token probability distribution produced by the \fmname backbone $g_{\theta}$ followed by an iBOT head $h^{\text{iBOT}}_{\theta}$ \cite{zhou2021ibot}.
In this objective, the student predicts masked patch tokens to match the teacher's unmasked outputs across seasonal views, encouraging temporally invariant representations and learning structural spatial dependencies (e.g., road connectivity and building density) that are essential for wealth estimation.
Concurrently, uniformity regularization promotes a well-distributed representation space and mitigates feature collapse, further enhancing the discriminative power of the learned features.

\noindent\textbf{Pretraining methods}.
\fmname model are pretrained for 800,000 steps on 4 H100 GPUS, with a batch size of 64 per GPU.
We use AdamW \cite{loshchilov2017decoupled} as the optimizer with a constant weight decay of 0.04.
We use a constant learning rate of $5\times 10^{-5}$ with a linear warm-up from zero for 10,000 steps.
To make model training memory-efficient, we adopt flash-attention 2 \cite{dao2023flashattention}, fully sharded data parallel \cite{zhao2023pytorch}, and bf16 mixed precision.
We use the drop path \cite{huang2016deep} with a drop probability of 0.3 as regularizer.

\noindent\textbf{Uncertainty-aware optimization}.
To adapt the pre-trained foundation model to wealth estimation tasks while accounting for label scarcity and noise, we introduce two complementary mechanisms:
1) Parameter-efficient adaptation. 
Instead of full fine-tuning, which is prone to overfitting on small survey datasets, we inject low-rank decomposition matrices into the query and value projection layers of the transformer blocks \cite{hu2022lora}.
For a pre-trained weight matrix $W_0 \in \mathbb{R}^{d \times k}$, the update is parameterized as $W_0 + \Delta W = W_0 + BA$, where $B \in \mathbb{R}^{d \times r}$ and $A \in \mathbb{R}^{r \times k}$ are trainable low-rank matrices ($r \ll d$). We freeze $W_0$ and optimize only $A$ and $B$;
2) Probabilistic regression head. 
Standard mean squared error (MSE) assumes deterministic labels and is suboptimal given inherent noise in survey data.
We therefore replace the deterministic head with a probabilistic head that predicts a Gaussian distribution $\mathcal{N}(\mu(x), \sigma^2(x))$ for the target wealth index $y$.
The model outputs both the predicted mean $\hat{y}$ and the heteroscedastic variance $\hat{\sigma}^2$. The optimization objective is Gaussian negative log-likelihood (GauNLL)\cite{nix1994estimating}:
\begin{equation}
\mathcal{L}_{\text{GauNLL}} = \log \hat{\sigma}^2 + \frac{(y - \hat{y})^2}{\hat{\sigma}^2}
\end{equation}
This objective naturally attenuates gradients from samples with high predictive uncertainty (likely noisy or jittered labels), thereby improving robustness to label noise.

\section{Evaluation Protocol}

\noindent\textbf{Metrics}.
We evaluate model performance using the coefficient of determination ($R^2$) and the squared Pearson correlation ($r^2$).
$R^2$ quantifies the fraction of AWI variance explained by the model and penalizes systematic bias in absolute magnitude, which is critical for calibrated wealth estimation in nowcasting, hindcasting, and continent-wide mapping.
In contrast, $r^2$ emphasizes the strength of linear association between predictions and observations, capturing how well the model preserves relative spatial and temporal ordering under cross-country and cross-year distribution shifts.
Reporting both metrics therefore provides a complementary assessment of performance in cross-domain settings: reliability in absolute level estimation and robustness in comparative ranking.

\noindent\textbf{Model selection for wealth mapping}.
We design a unified spatiotemporal \textit{k}-fold cross-validation strategy ($k=5$) to evaluate five generalization scenarios.
Data are grouped by country and year, then split into spatially disjoint folds within each country-year entry (illustrated in Supplementary Fig.~1).
Let $C=\{c_1, c_2, ..., c_n\}$ denote the country set ($n=34$), with target country $A=\{c_i\}$, other countries $B=C - A$, target year $T_1$, and all other years $T_2$.
We evaluate five different aspects of generalization:
(1) \textit{Out-of-country}.
We train on a subset of $B$, validate on the remaining subset of $B$, and test on $A$, which evaluates out-of-country generalization.
(2) \textit{In-country-and-out-of-year}.
We train on four folds of $A$ from $T_2$, validate on the remaining fold $i$, and test on fold $i$ in $T_1$, which evaluates out-of-year generalization.
(3) \textit{In-country-and-in-year}.
For target country $A$, we train on data from $T_2$ and three folds from $T_1$, validate on fold $i$ from $T_1$, and test on fold $(i+1)\bmod k$ from $T_1$, thereby evaluating generalization when in-year data are available.
(4) \textit{All-countries-and-out-of-year}.
We train on $B$ and four folds of $A$ from $T_2$, validate on the remaining fold $i$ of $A$ from $T_2$, and test on fold $i$ in $T_1$, thereby evaluating whether cross-country priors improve in-country out-of-year generalization.
(5) \textit{All-countries-and-in-year}.
We train on $B$, on $A$ from $T_2$, and on three folds of $A$ from $T_1$, validate on fold $i$ from $T_1$, and test on fold $(i+1)\bmod k$ from $T_1$, thereby evaluating generalization when all data are available. 
Based on this strategy, we totally trained 1,715 models to produce the analysis in \ed Fig.~\ref{exfig:5}.

\section{Model inference}
We developed a fast and efficient continent-scale mapping pipeline by using multithreading to parallelize data acquisition, with each CPU node assigned to a single country. This design enables parallel downloading both within and across countries. In parallel, \fmname performs inference on GPUs, allowing prediction to proceed concurrently with data acquisition on a separate hardware architecture. We further scale inference by distributing the model across multiple NVIDIA A100 GPU nodes, each responsible for one country. Using this pipeline, we mapped the entire African continent in approximately 3 hours at a total cost of 36.5 GPU-hours.

\section{Data availability}

Most data used in this study are publicly available, except for census data provided by the World Bank Group (\url{https://www.worldbank.org/}).
Landsat satellite imagery is available from SSL4EO \cite{stewart2023ssl4eo} and NASA (\url{https://science.nasa.gov/mission/landsat/open-data/}).
DHS data are available from the Demographic and Health Surveys Program (\url{https://www.dhsprogram.com/}).
IHS data are available from the International Household Survey Network (\url{https://www.ihsn.org/}).
Population data are obtained from the 2020 WorldPop dataset (\url{https://www.worldpop.org/}).
All plots were generated using Matplotlib \cite{Hunter:2007} and GeoPandas \cite{kelsey_jordahl_2020_3946761}.
Our continent-wide maps of wealth and wealth changes for Africa is accessible via Google Earth Engine (\url{https://zhuozheng2017.users.earthengine.app/view/tempov-africa-wealth-watch-2015-2025}).

\clearpage
\setcounter{figure}{0}
\captionsetup[figure]{name=Extended Data Fig., labelsep=pipe}

\begin{figure*}[h]
\centering
\includegraphics{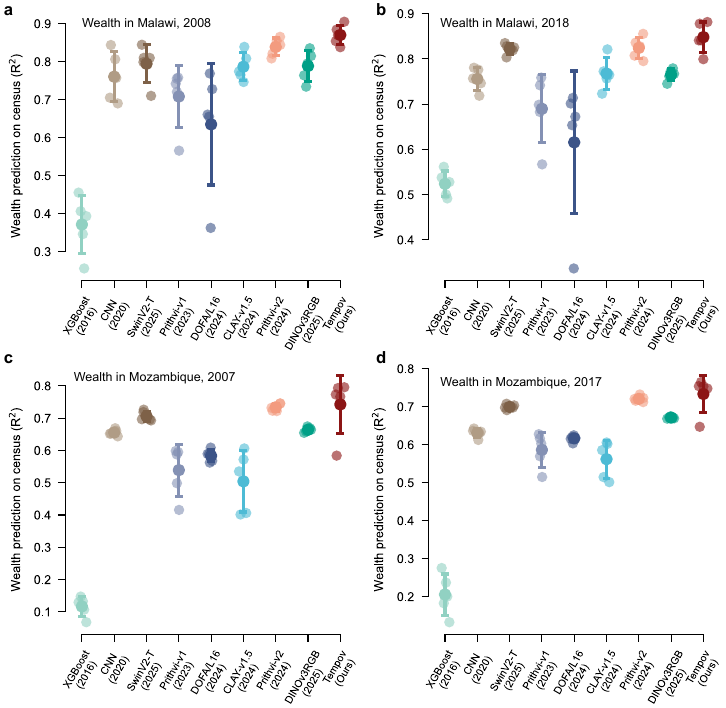}
\caption{Benchmarking wealth level prediction capabilities in data-rich scenario where the model is trained on full census data.
Performance comparison (coefficient of determination $R^2$) across Malawi (2008 \textbf{a}, 2018 \textbf{b}) and Mozambique (2007 \textbf{c}, 2017 \textbf{d}) for wealth level prediction. 
The proposed method is evaluated against XGBoost, CNN, SwinV2-T and five geospatial foundation models. 
Data points represent independent validation folds ($n=5$); center dots and error bars indicate mean $\pm$ 1 s.d.
}
\label{exfig:1}
\end{figure*}

\begin{figure*}[h]
\centering
\includegraphics[width=0.95\textwidth]{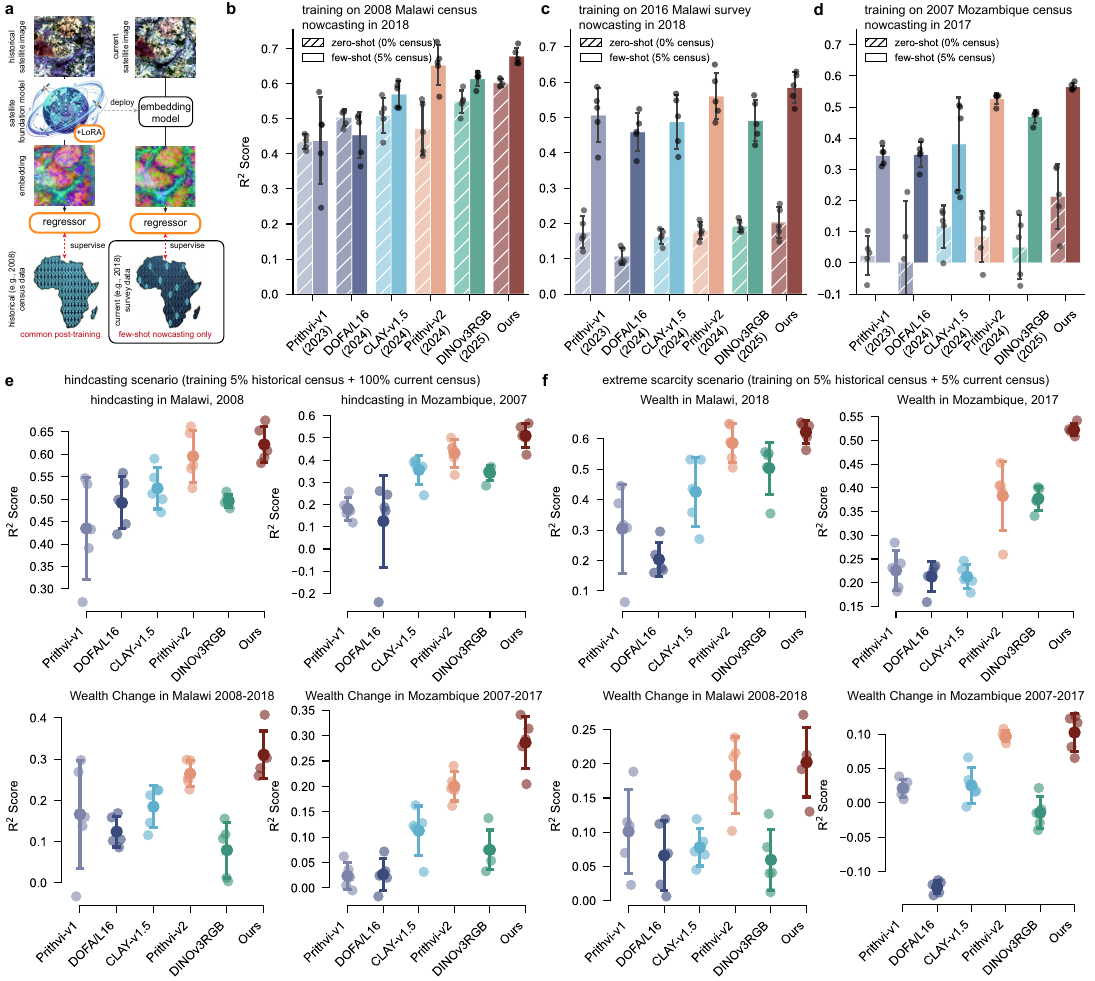}
\caption{Versatile spatiotemporal adaptation capabilities across varying data regimes.
\textbf{a}, Schematic of the two-stage adaptation framework. 
The satellite foundation model initially undergoes common post-training using historical priors (census or survey) via Low-Rank Adaptation (LoRA). 
It is then deployed for predict on the given imagery in either a zero-shot (direct transfer) or few-shot (supervised by limited data) setting.
\textbf{b-d}, Performance evaluation ($R^2$) of wealth nowcasting in Malawi (b, c) and Mozambique (d). 
Models trained on historical census or survey labels (Malawi 2008 (b); Malawi 2016 (c); Mozambique 2007 (d)) are evaluated on ground truth data from a decade later (2018 and 2017, respectively).
\textbf{e}, Retrospective hindcasting. Assessment of models trained on current data (100\%) supplemented with sparse historical labels (5\%) to reconstruct past wealth states and track decadal changes.
\textbf{f}, Benchmarking under extreme scarcity. Performance of wealth and change estimation when supervision is restricted to minimal subsets (5\% historical + 5\% current labels).
In all panels, ``Ours'' denotes the proposed method; data points represent $n=5$ independent folds, and error bars indicate mean $\pm$ 1 s.d.
}\label{exfig:3}
\end{figure*}

\begin{figure*}[h]
\centering
\includegraphics[width=0.95\textwidth]{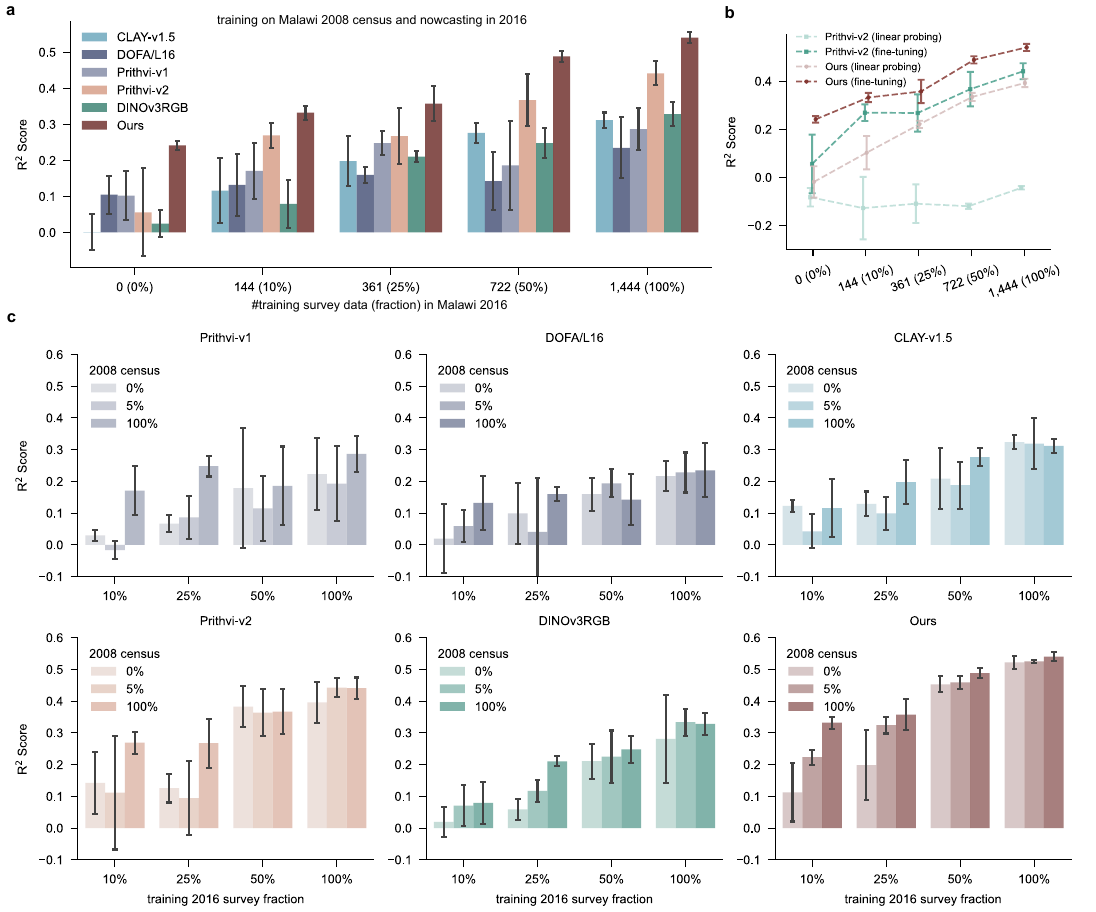}
\caption{Synergistic scaling of historical census priors and current survey data. 
\textbf{a}, Impact of increasing survey sample size on nowcasting accuracy. 
Models are adapted using historical 2008 census data and real-world household 2016 survey data from Malawi, ranging from 0\% to 100\% coverage. 
\textbf{b}, Comparison between the proposed model (Ours) and Prithvi-v2 using fine-tuning and linear probing strategies across the training data fraction as in \textbf{a}.
\textbf{c}, Detailed breakdown of data efficiency. 
Performance is benchmarked across varying regimes of data availability: historical census data (0–100\%, represented by color) and current survey labels (10–100\%, x-axis). 
The results highlight that while baseline models struggle with limited survey data, the proposed framework robustly integrates historical knowledge to compensate for the sparsity of real-world labels. Error bars represent the standard deviation across multiple runs.
}\label{exfig:4}
\end{figure*}

\begin{figure*}[h]
\centering
\includegraphics{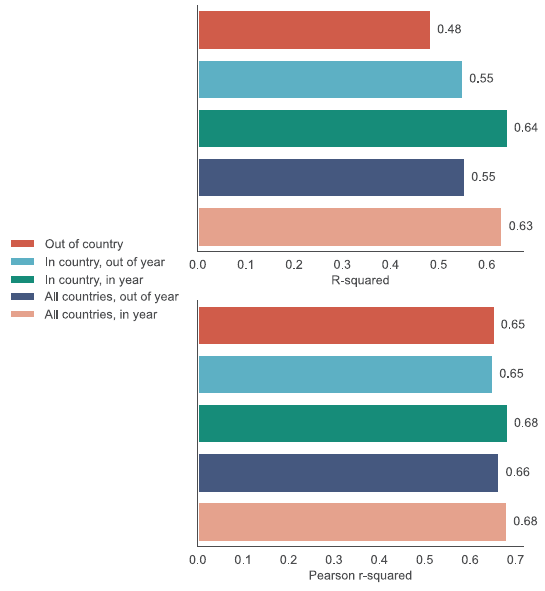}
\caption{Performance comparison of five spatial and temporal generalization strategies for model selection, evaluated by coefficient of determination ($R^2$, top) and Pearson correlation squared ($r^2$, bottom).
}\label{exfig:5}
\end{figure*}

\begin{figure*}[h]
\centering
\includegraphics[width=0.9\textwidth]{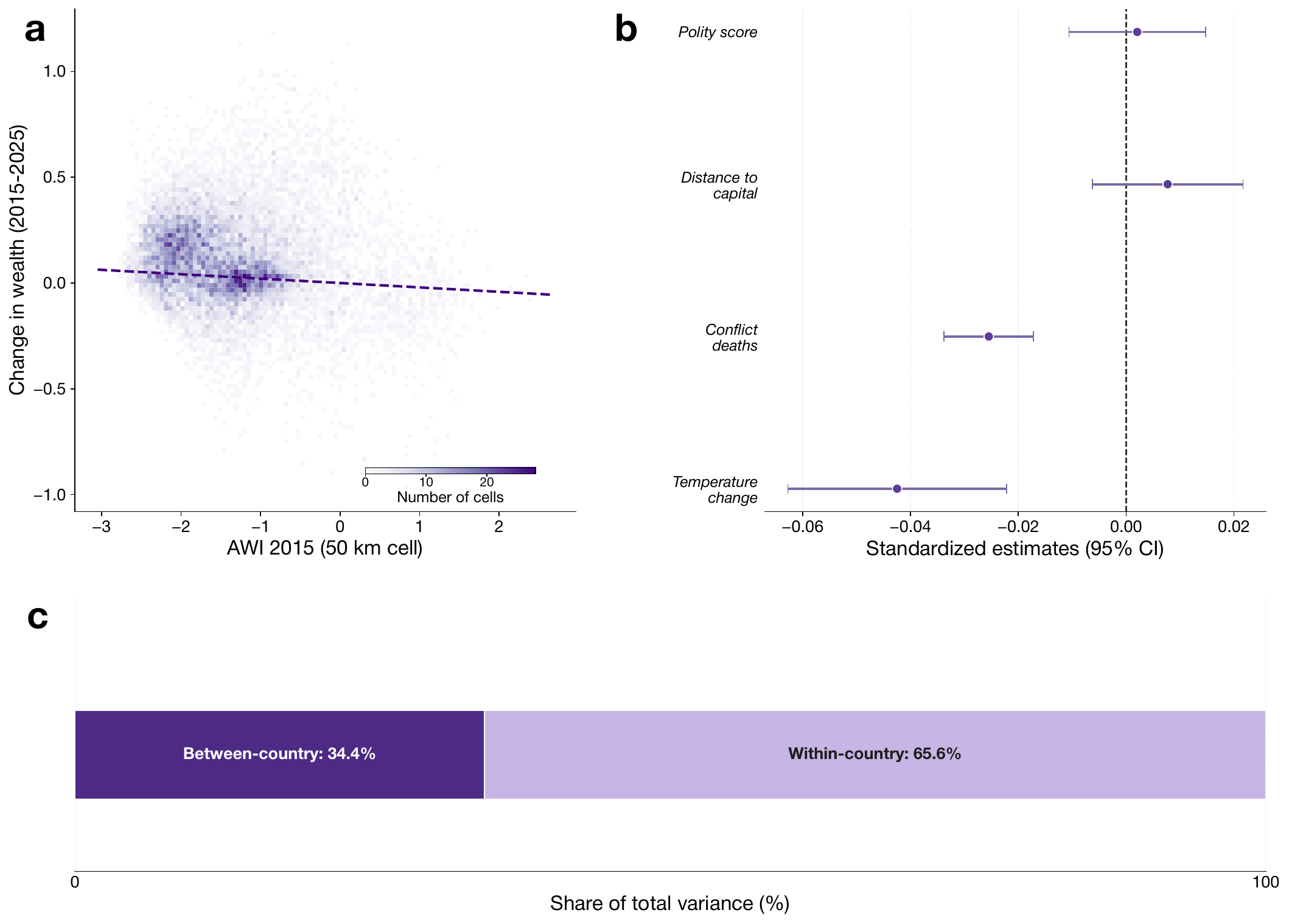}
\caption{Determinants of AWI change (2015--2025). We evaluate whether \fmname AWI predictions capture growth dynamics and their determinants. \textbf{a.} Statistically significant $\beta$-convergence across 50-km grid cells covering 53 African countries ($\hat{\beta} = -0.021$; 95\% CI: $[-0.024, -0.017]$). Shade in plot captures the density of the data. \textbf{b.} Aggregate contributions of established growth determinants: institutional quality (mean Polity V scores by country), remoteness (distance from each grid centroid to its respective national capital), conflict intensity (total casualties within each 50-km grid cell from the UCDP geo-referenced conflict events dataset), and environmental factors (ERA5 2-meter temperature change, computed using grid-specific linear trends). All variables cover the 2015--2025 study period. Estimates are population-weighted using 2015 WorldPop gridded data at 50-km resolution; inputs are standardized to express changes in standard-deviation units. \textbf{c.} Variance decomposition of growth into local and country-level components.}
\label{exfig:6}
\end{figure*}

\clearpage
\onecolumn
\setcounter{page}{1}

\begin{center}
    \LARGE \textbf{Supplementary Information}
\end{center}

\setcounter{figure}{0}
\setcounter{table}{0}
\captionsetup[figure]{name=Supplementary Fig., labelsep=pipe}
\captionsetup[table]{name=Supplementary Table, labelsep=pipe}

\setcounter{section}{0}
\renewcommand{\thesection}{}

\normalsize\textbf{Table of contents}

\startcontents[sections]
\printcontents[sections]{}{1}{\setcounter{tocdepth}{2}}

\clearpage

\section{A. Tempov model architecture}

In our pretraining setting, the backbone is a vision transformer with a patch size of 16. 
The model operates on 6-channel inputs and employs a convolutional patch embedding layer with kernel size and stride both set to 16 to produce 1024-dimensional token embeddings. 
A learnable class token and four learnable storage tokens are prepended to the patch sequence, while a learnable mask token is used for masked token replacement during self-supervised learning. 
The encoder consists of 24 pre-normalized Transformer blocks with 16 attention heads per block. 
Each block follows a standard pre-norm design, where LayerNorm \cite{ba2016layer} is applied before multi-head self-attention \cite{attn} and before the feed-forward subnetwork, and each sublayer is wrapped with a residual connection scaled by LayerScale initialized to 1e-5. 
The feed-forward branch is implemented as a two-layer MLP with GELU \cite{hendrycks2016gaussian} activation and an intermediate dimension of 4096. 
Positional information is injected through 2D rotary positional encoding (RoPE) \cite{rope} rather than learnable absolute positional embeddings, with a RoPE base of 100, separate normalization for horizontal and vertical coordinates.
In addition, stochastic depth is applied with a drop-path rate of 0.3 during pretraining.
The encoder outputs the normalized class token as the global image representation, together with normalized patch tokens as the local yet dense image representation.

\section{B. All-African DHS data and satellite imagery for continent-wide wealth mapping}

We collect and preprocess DHS data from nearly all African countries, covering 34 countries and 89 country-year entries.
This yields 42,993 clusters, each paired with a Landsat image and an AWI value.
For wealth mapping, we use median-composite Landsat imagery from 2024--2026 as 2025 baseline imagery; each image covers a 6\,km $\times$ 6\,km area to accommodate DHS GPS coordinate jitter.

\begin{table}[htbp]
\centering
\caption{DHS data we used for continent-wide wealth mapping}
\label{tab:country_year_list}
\begin{tabular}{ll}
\toprule
\textbf{Country} & \textbf{Years} \\
\midrule
Angola & 2011 \\
Benin & 2012, 2017 \\
Burkina Faso & 2010, 2014, 2017 \\
Burundi & 2010, 2012, 2016 \\
Cameroon & 2011, 2018, 2022 \\
Chad & 2014 \\
Comoros & 2012 \\
Congo & 2013, 2023 \\
Côte d'Ivoire & 2012, 2021 \\
Ethiopia & 2011, 2016, 2019 \\
Gabon & 2012, 2019 \\
Gambia & 2019 \\
Ghana & 2014, 2016, 2019, 2022 \\
Guinea & 2018, 2021 \\
Kenya & 2014, 2015, 2020, 2022 \\
Lesotho & 2014, 2023 \\
Liberia & 2011, 2013, 2016, 2019, 2022 \\
Madagascar & 2011, 2013, 2016, 2021 \\
Malawi & 2010, 2015, 2017 \\
Mali & 2012, 2015, 2023 \\
Mauritania & 2020 \\
Mozambique & 2011, 2015, 2018, 2022 \\
Namibia & 2013 \\
Niger & 2012, 2021 \\
Nigeria & 2013, 2015, 2018, 2021, 2024 \\
Rwanda & 2010, 2015, 2019 \\
Senegal & 2020, 2023 \\
Sierra Leone & 2013, 2016, 2019 \\
South Africa & 2016 \\
Tanzania & 2010, 2012, 2015, 2017, 2022 \\
Togo & 2013, 2017 \\
Uganda & 2011, 2014, 2016, 2018 \\
Zambia & 2013, 2018, 2024 \\
Zimbabwe & 2010, 2015 \\
\bottomrule
\end{tabular}
\end{table}

\clearpage

\section{C. Synergizing historical census with sparse surveys}

The nowcasting scenario with historical data and current sparse surveys is a typical real-world setting in which up-to-date wealth measurements are required despite severely limited contemporary household surveys.
We systematically investigated the data efficiency of our framework by scaling the available current survey samples from 0\% to 100\% (\ed Fig.~\ref{exfig:4}a).
The results reveal a striking advantage in data-scarce regimes: our model achieves a competitive $R^2$ score with only 10\% of the survey data (144 image-AWI samples), matching the performance that baseline foundation models (e.g., CLAY-v1.5, Prithvi v1) attain only with 100\% of the data.
This implies an order-of-magnitude reduction in the cost of household survey collection, making high-frequency monitoring feasible for resource-constrained organizations.

To understand the optimal adaptation mechanism, we compared LoRA fine-tuning against linear probing (\ed Fig.~\ref{exfig:4}b).
Consistently, fine-tuning (darker dashed lines) outperforms linear probing (lighter dashed lines) across all data regimes for top-2 foundation models.
This confirms that deep parameter adaptation is essential to fully bridge the domain gap and realign the model’s semantic representations with the specific socioeconomic context of the target year.
This result suggests that linear models built upon pre-computed model embeddings, such as those now released by several foundation models, may have limited performance for wealth mapping. 
At the same time, our model’s linear probing trajectory (light brown line) does improve with sample size, reaching an average $R^2$ of 39\%, which even exceeds the performance of Prithvi-v2 fine-tuned on 50\% of survey data (average $R^2$ of 36\%).
This suggests that our foundation model learns representations with a strong linear relationship to wealth indicators directly during pre-training, whereas Prithvi-v2 does not show this behavior, with linear probing performance remaining near or below zero regardless of data availability.

To further probe the interplay between historical census and current surveys, we systematically varied the fraction of census and survey data used in training (\ed Fig.~\ref{exfig:4}c).
We observe a strong compensatory synergy: the benefit of incorporating historical census data (comparing light vs. dark bars) is most pronounced when current survey data is scarce (e.g., at the 10\% fraction). 
In this low-data regime, our model effectively ``retrieves'' structural knowledge from the past to stabilize predictions, yielding a massive performance boost. 
In contrast, baseline models (e.g., Prithvi-v2, DOFA) show limited sensitivity to historical priors (flatter bar gradients), suggesting an inability to effectively transfer past knowledge. 
As current data becomes abundant ($\geq$50\%), the reliance on historical priors naturally diminishes, confirming that our framework dynamically balances past structural insights with present-day evidence.

\clearpage

\section{D. Temporal patterns between asset wealth and gpd per capita}

\begin{figure*}[h]
\centering
\includegraphics[width=0.92\textwidth]{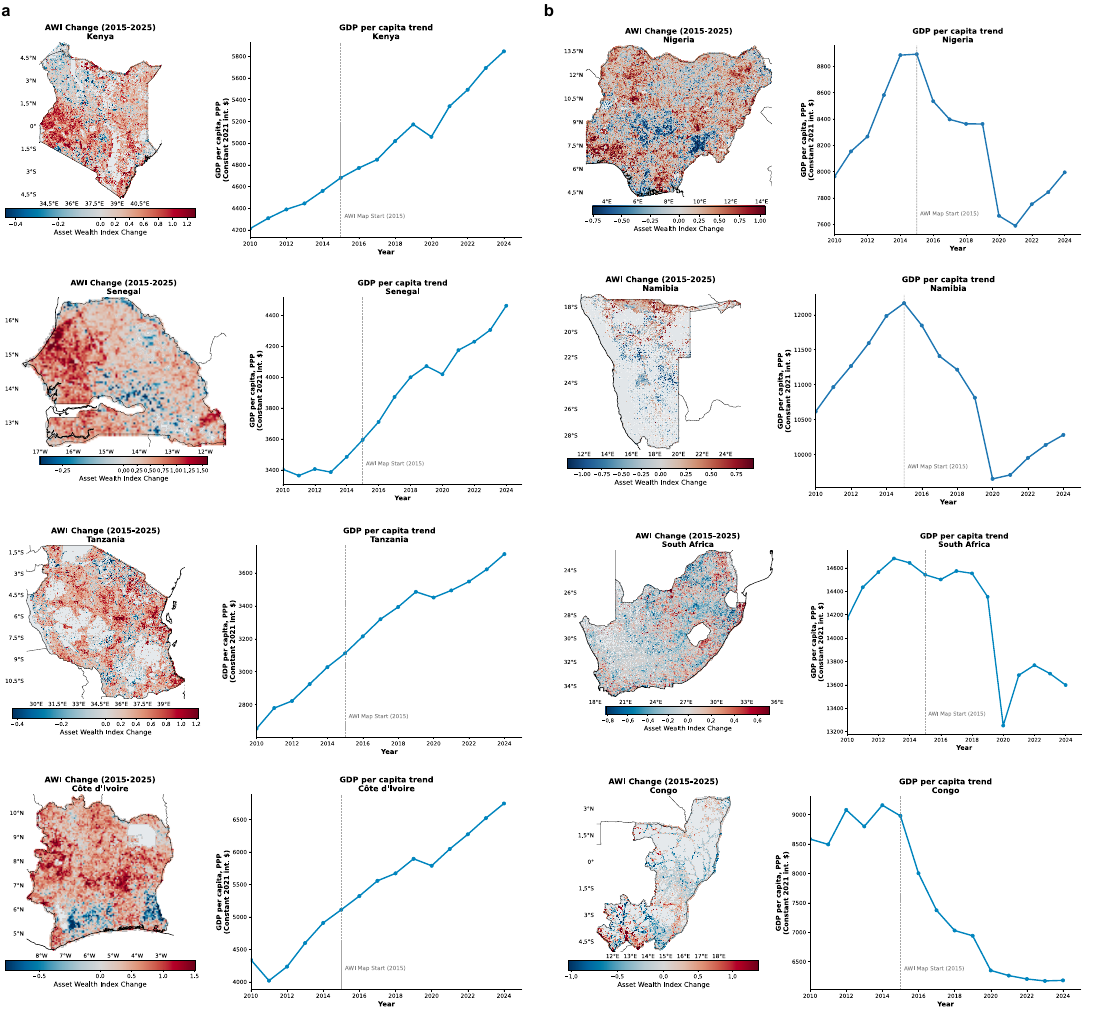}
\caption{Comparison of predicted asset wealth change with national GDP per capita trends across selected African countries.
\textbf{a}, Countries with broadly increasing GDP per capita over the past decade, including Kenya, Senegal, Tanzania and Côte d’Ivoire. 
\textbf{b}, Countries with declining GDP per capita trajectories, including Nigeria, Namibia, South Africa and Congo. 
For each country, the left panel shows the predicted change in the Asset Wealth Index (AWI) from 2015 to 2025, with red indicating relative increases and blue indicating relative declines, and the right panel shows the national GDP per capita trend in purchasing power parity terms (constant 2021 international dollars) from World Bank data (\url{https://data.worldbank.org/indicator/NY.GDP.PCAP.PP.KD}). 
The vertical dashed line marks 2015, the baseline year of the AWI maps. 
The AWI changes are generally consistent with broad macroeconomic trends. 
The maps also reveal substantial within-country spatial heterogeneity that is not captured by national GDP per capita alone.
}
\label{supp:fig:1}
\end{figure*}

\begin{figure*}[h]
\centering
\includegraphics[width=0.9\textwidth]{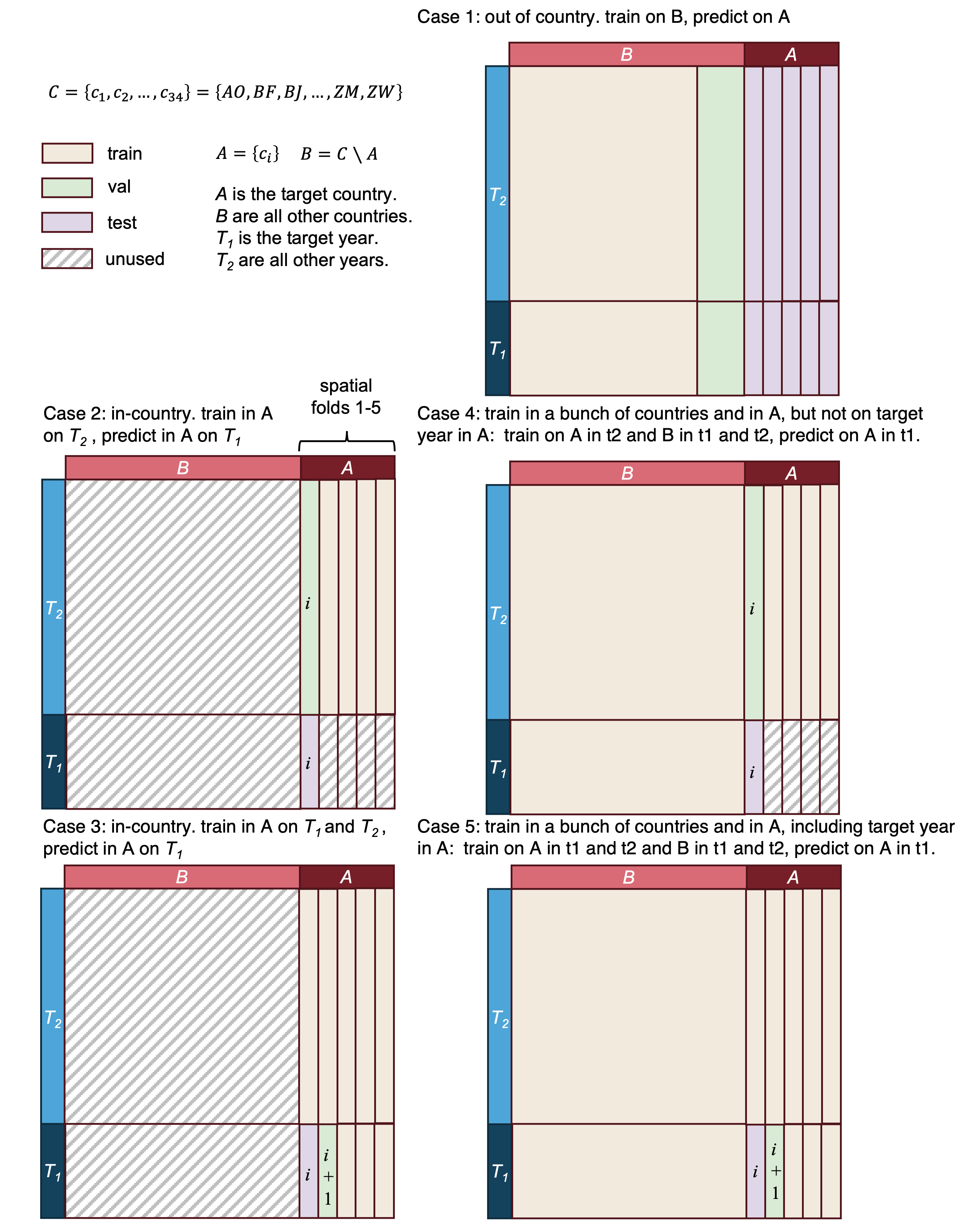}
\caption{Unified spatiotemporal \textit{k}-fold cross-validation strategy.
}
\label{supp:fig:2}
\end{figure*}


\end{document}